THE EUROPEAN
PHYSICAL JOURNAL C

# Charged particle dynamics in the surrounding of Schwarzschild anti-de Sitter black hole with topological defect immersed in an external magnetic field

**Sidra Shafiq**[1], **Saqib Hussain**[2,a], **Muhammad Ozair**[3], **Adnan Aslam**[4,b], **Takasar Hussain**[3]

[1] Department of Physics, Government College University (GCU), Lahore, Pakistan
[2] Institute of Astronomy, Geophysics and Atmospheric Sciences (IAG), University of São Paulo (USP), São Paulo, Brazil
[3] Department of Mathematics, COMSATS University, Attock Campus, Attock, Pakistan
[4] School of Electrical Engineering and Computer Science (SEECS), National University of Sciences and Technology (NUST), Islamabad, Pakistan

**Abstract** In this paper, geodesic motion of the charged particles in the vicinity of event horizon of Schwarzschild anti-de-Sitter black hole (BH) with topological defects has been investigated. Weakly magnetized environment is considered in the surrounding of BH which only effects the motion of the particles and doesn't effect the geometry of the BH. Hence, particles are under the influence of gravity and electromagnetic forces. We have explored the effect of magnetic field on the trajectories of the particles and more importantly on the position of the innermost stable circular orbit. It is observed that the trajectories of the particles in the surrounding of BH are chaotic. Escape conditions of the particles under the influence of gravitomagnetic force are also discussed. Moreover, the escape velocity of particles and its different features have been investigated in the presence and absence of magnetic field. Effect of dark energy on the size of event horizon, mass of the BH and stability of the orbits of the particles have also been explored in detail. These studies can be used to estimate the power of relativistic jets originated from the vicinity of BH.

## 1 Introduction

Universe is in accelerating phase, was first confirmed by the Supernova of type Ia, large scale structure [1] and quasars which are the most luminous persistent sources in the universe, observed up to redshifts of $z \approx 7$ [2]. Quasars can be used to study the cosmology of our universe in a relatively unexplored adolescence. There is an increase in the dark energy (DE) density if one consider an evolution of the equation of state (EoS) of DE [3]. If this model is right, DE may be in a particularly virulent and implausible form called phantom energy [4]. There are different models proposed to understand the nature of DE in our universe [5]. So far, several scalar field DE models have been studied such as quintessence [6], phantom [4], K-essence [7], tachyon [8] and quintom [9]. The Big Rip scenario [10] could justify the DE, that such energy is called phantom energy.

The nature of the EoS of phantom energy is supernegative which leads to the growing energy density of phantom energy, $\rho_d \sim a^{-3(1+\omega)}$ for a parameter $\omega < -1$.

In the framework of inflationary paradigm, recent cosmological observations indicates that a very small relict vacuum energy ($\lambda = 8\pi G/c^2(\rho_{\text{vac}}) > 0$), or, generally, a DE demonstrating repulsive gravitational effect, has to be introduced to explain dynamics of the expanding Universe [11,12]. An effective influence of the repulsive cosmological constant has been demonstrated for many astrophysical situations related to active galactic nuclei and their central BH [13]. In spacetimes with a repulsive cosmological term $\lambda$, the motion of photons was extensively investigated [14] and the dynamics of massive test particles is also studied [15]. The role of the $\lambda$ can be significant for both the geometrically thin Keplerian accretion discs [13] and the toroidal discs [16] orbiting supermassive BH (Kerr superspinars) in the central parts of giant galaxies. Both high-frequency quasi-periodic oscillations and jets originating at the accretion discs can be reflected by current carrying string loops in S-AdS and Kerr-AdS spacetimes [17]. According to [18], the accretion of a phantom test fluid onto a Schwarzschild BH will induce the mass of the BH to decrease, however the back-reaction was ignored in their calculation. Using new exact solutions describing BH in a background Friedmann-Robertson-Walker universe, it is found that the physical BH

[a] e-mail: s.hussain2907@gmail.com (corresponding author)
[b] e-mail: adnan.aslam@seecs.edu.pk

Springer

mass increases due to the accretion of phantom energy [19]. The effects of phantom accretion on the charged BH in 4D have been explored [20]. They pointed out that if mass of BH becomes smaller (due to accretion of phantom energy) than its charge, then the Cosmic Censorship Hypothesis is violated. Masses of all BH tend to zero in the phantom energy universe approaching to the Big Rip [18].

There are many models which explain the expansion of the universe, dynamics of photons and accretion discs with respect to the growth of DE density. Mass of the BH can increase or decrease due to accretion of phantom energy. These studies motivate us to explore the geodesic motion of a charged particle in the vicinity of S-AdS-BH with topological defect in the presence of magnetic field. High energy phenomenon occurring near the BH such as formation of jets and accretion disks can be elaborated with this study. Change in the BH mass can be explored with this study as well. Due to the presence of strong gravitational and electromagnetic fields, charged particles in general do not follow stable orbits and inter-particle collisions are inevitable. Magnetic field present in the nearby surrounding of the BH [21] supports the large scale jets. These jets are most likely the source of galactic and extragalactic cosmic rays. The magnetic field is homogeneous at infinity. This field strongly effects the dynamics of the particles and location of their innermost stable circular orbits (ISCO) around BH. Therefore, during the motion of a particle in the vicinity the BH, it is under the influence of DE, gravitational and electromagnetic forces.

In the present article, it is considered that a charged particle is orbiting in the ISCO of the S-Ads-BH and is hit by a radially incoming particle. The aftermath of a collision will depend on the energy of the incoming particle which may result in one of the three outcomes: the charged particle may escape to infinity, be captured by the BH or keeps orbiting in ISCO. However, predicting the nature of outcome is compounded by the facts that the particle is charged and interacts with the magnetic field.

The outline of the paper is as follows: In Sect. 2 we will discuss magnetic field strength and calculate its expression in the surrounding of S-Ads-BH. In Sect. 3 we will present the detailed mathematical derivation of our model to study the dynamics of the charged particle in the vicinity of S-Ads-BH along with the properties of ISCO, e. g. effective potential, escape velocity and DE effects on the event horizon. The discussion and the results are given in the last section.

## 2 Magnetized environment of the BH

It is very interesting to study the electromagnetic fields and particle motion in the surrounding of BH with the aim to get new tool and physical insights for studying new important relativistic effects. This kind of demonstration and research can be interesting because of the existence of both theoretical and experimental evidences that a magnetic field must be present in the vicinity of BH. Here we use the weak magnetic field limit such that the energy and momentum of this field can not change the background geometry of BH [29]. The strength of magnetic field in the surrounding of a BH with mass $M$ should satisfy the condition given in [29],

$$\mathcal{B} \ll \mathcal{B}_{max} \sim 10^{19} \frac{M_\odot}{M_{BH}} Gauss. \quad (1)$$

The maximum magnetic field strength in the universe is found in the magnetars and their high magnetic field strength can effect the surrounding space time geometry. To study the dynamics of magnetars one has to consider the high resolution megneto-hydro-dynamics (MHD) simulations. But, here we are dealing with stellar mass and supermassive BHs. The magnetic field strength on the BH horizon is $\mathcal{B} \sim 10^8$ G for stellar mass BH and $\mathcal{B} \sim 10^4$ G for the supermassive BH as given in [31]. So, the above condition satisfies for stellar BH as well as the supermassive BH.

The BHs which satisfy the condition (1) are known as "weakly magnetized" [29] provided that the magnetic field strength lies within $\mathcal{B} \sim (10^4 - 10^8)$ Gauss $\ll 10^{19}$ Gauss. The magnetic field is responsible for transferring energy to the particles moving in the surrounding of BH so that their escape to spatial infinity is possible [29]. Hence, the collision of charged particles near the BH may produce much higher energy in the presence of magnetic field than its absence. The effects of magnetic fields on the charged particles moving around a BH are considerably strong as studied in [30,32,33]. The presence of axially symmetric magnetic field makes the timelike geodesics in the modified gravity BH very chaotic [34].

We are investigating the motion of a charged particle as it is effected by both magnetic field in the BH exterior and gravitational field. We are using the same magnetic field profile as it is used in [23,35] The general Killing vector equation is [26]

$$\Box \xi^\mu = 0, \quad (2)$$

where $\xi^\mu$ is a Killing vector. The 4-potential and magnetic field vector is given as [23,35]

$$A^\mu = \frac{\mathcal{B}}{2} \xi^\mu_{(\phi)}, \quad \mathcal{B}^\mu = -\frac{1}{2} e^{\mu\nu\lambda\sigma} F_{\lambda\sigma} u_\nu, \quad (3)$$

where $\mathcal{B}^\mu$ is the magnetic field tensor and $F_{\lambda\sigma}$ is the Maxwell tensor. Using Eqs. (2) and (3) we have obtained the magnetic field given below,

$$\mathcal{B}^\mu = \mathcal{B} \frac{1}{\sqrt{f(r)}} \left[ \cos\theta \delta^\mu_r + \frac{\sin\theta \delta^\mu_\theta}{r} \right]. \quad (4)$$

Here we have considered magnetic field to be directed along the vertical (z-axis) and $\mathcal{B} > 0$. This magnetic field configuration is similar to [23,35] with the change of constants of motion (Energy $\varepsilon$ and angular momentum $L$) corresponding to the considered geometry. In Fig. 1 we plot the magnetic field in the dimensionless form $b = \mathcal{B}M$ against $r = r/M$ using Eq. (4) for equatorial plane ($\theta = \pi/2$). We have plotted all the quantities in dimensionless form in all the figures. Here we have considered a weakly magnetized environment which is followed by an external static axisymmetric magnetic field. Moreover, this field is uniform at spatial infinity and blows up only at the horizon and $r = 0$.

## 3 Particle dynamics in the surrounding of magnetized BH

### 3.1 Schwarzschild Anti-de Sitter BH geometry

The Schwarzschild AdS (S-AdS) BH with topological defect is described as [24]

$$ds^2 = f(r)dt^2 - \frac{dr^2}{f(r)} - r^2(d\theta^2 + \sin^2\theta d\phi^2),$$
$$f(r) = 1 - \eta^2 - \frac{2M}{r} + \frac{r^2}{a^2}. \quad (5)$$

As $f(r)$ is a cubic polynomial in $r$, we have three root of $f(r) = 0$; one is real and two are imaginary. Therefore, real root is the event Horizon (EH) of the BH given by;

$$r_+ = r_h = \frac{A^2 - 3a^2(1-\eta^2)}{3A}$$
$$A = 3a\left[\frac{M}{a} + \sqrt{\frac{1}{27} + \frac{M^2}{a^2} - \frac{\eta^2}{9}\left(1 - \eta^2 + \frac{\eta^4}{3}\right)}\right], \quad (6)$$

$$r_{im1} = \frac{2(-3)^{2/3}a^2(\eta^2-1) + \sqrt[3]{3}\left(-1 - i\sqrt{3}\right)\left(9a^2M + \sqrt{3}\sqrt{27a^4M^2 - a^6(\eta^2-1)^3}\right)^{2/3}}{6\sqrt[3]{9a^2M + \sqrt{3}\sqrt{27a^4M^2 - a^6(\eta^2-1)^3}}}, \quad (7)$$

$$r_{im2} = \frac{i\left(\sqrt{3}+i\right)\sqrt[3]{9a^2M + \sqrt{3}\sqrt{27a^4M^2 - a^6(\eta^2-1)^3}}}{2\cdot 3^{2/3}}$$
$$- \frac{\sqrt[3]{-\frac{1}{3}}a^2(\eta^2-1)}{\sqrt[3]{9a^2M + \sqrt{3}\sqrt{27a^4M^2 - a^6(\eta^2-1)^3}}}. \quad (8)$$

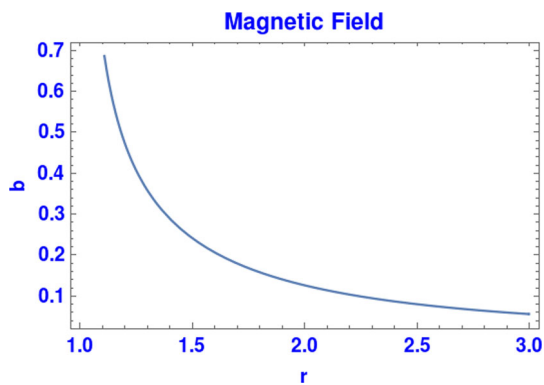

**Fig. 1** This plot shows trend of the magnetic field (b) in the surrounding of weakly magnetized BH. It is plotted for $\eta = 10^{-6}$ and $a = 0.5$. These values are followed from the BH metric given by Eq. (5)

Where $M$ is the mass of the BH, $\eta$ is the parameter that is related to the scale of symmetry breaking (topological defect) and $a$ is related with cosmological constant as $a^2 = 3/\lambda$ [38]. So for a typical grand unification scale $\eta \approx 10^{-6}$ it can be written as $1 - \eta^2 \approx 1$. Moreover, the parameter $a$ is not the spin of BH. We assume a magnetic field around the S-Ads BH to be uniform at infinity. The S-Ads space-time is static and spherically symmetric, which implies the existence of two commuting Killing vector fields, $\xi^\mu_{(t)} = (1,0,0,0)$ and $\xi^\mu_{(\phi)} = (0,0,0,1)$.

### 3.2 Horizon and dark energy

The EH depends on parameter $a$; the DE parameter and the symmetry breaking parameter $\eta$; of the BH. The radius of EH increases with the increase of $\eta$ and decreases with the increase of $a$; as shown in Fig. 2. As $\eta$ increases the size of the BH-EH increases. The parameter $a$ is inversely proportional to DE parameter $\lambda$, so large value of $a$ means less DE. In spherical symmetric space-time like Schwarzschild-BH, EH and apparent horizon (AH) coincides. Thus AH increase with the increase of DE while that of the cosmic AH decreases, during the evolution of the universe it may lead to a Big Rip [19]. There exists a moment at which the BH-AH coincides with the cosmological AH, and after that the BH singularity becomes naked which is forbidden by the Cosmic Censorship Conjecture.

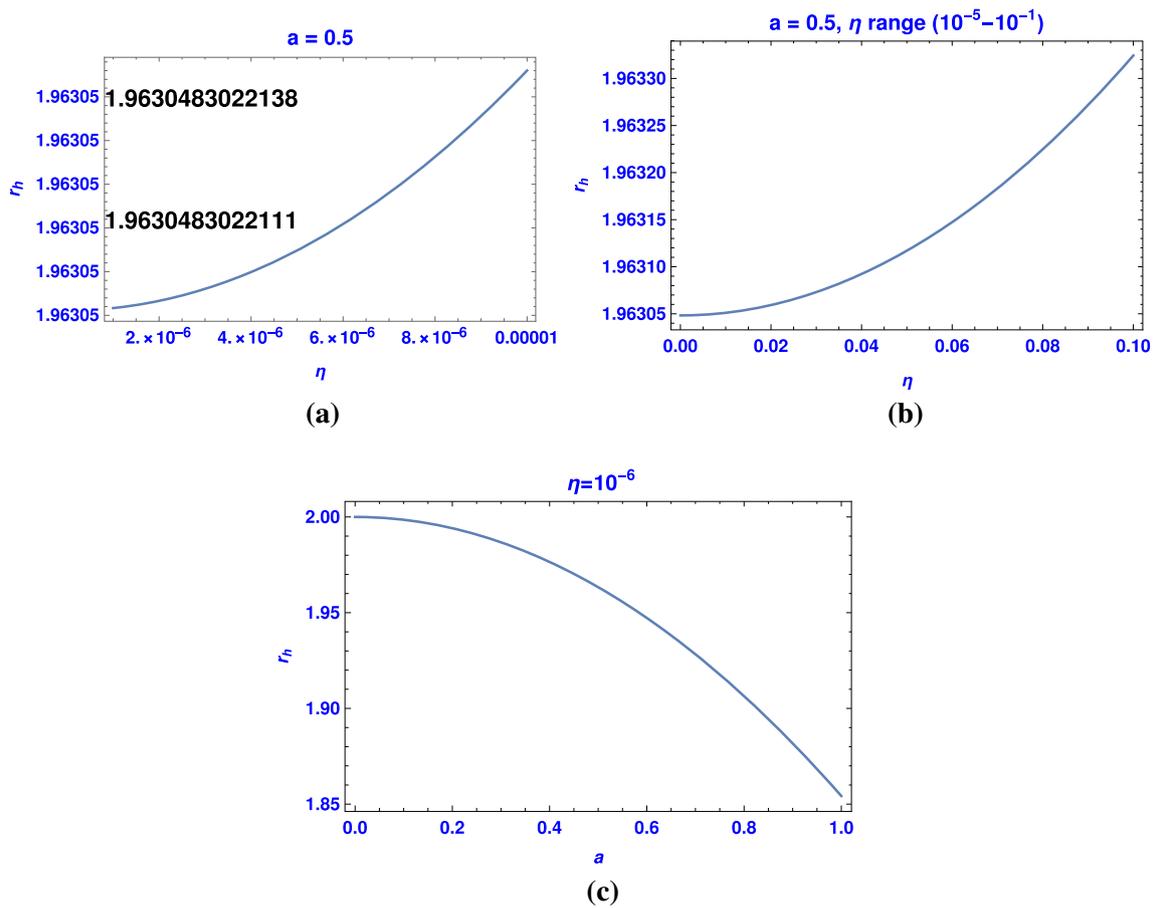

**Fig. 2** The behaviour of horizon as a fuction of parameters $\eta$ and $a$ of the BH is plotted. In **a** ordinate is explicitly mentioned in black color as it change only after 12th decimal place for the small range of $\eta \sim 10^{-6} - 10^{-5}$. In all the panels $M$ is set to unity

The parameter of topological defect $\eta$ has very small value $\approx 10^{-6}$ as given in [24]. We plotted the horizon against $\eta$ for different ranges as shown in Fig. 2. The horizon depends only on $\eta^2$ as it appears in Eq. (6). We plotted $r_h$ for a very small range of $\eta \sim (10^{-6} - 10^{-5})$ for which the increase in horizon is very small, it only appears after the 12th decimal place as shown in panel (a) of Fig. 2 (numbers in black color). Therefore, in panel (a) of Fig. 2 there is only one number (1.96305) appearing on the ordinate. However, we have plotted the $r_h$ for large range of $\eta$ in panel (b) of Fig. 2, where the increase in horizon is more visible, comparatively.

### 3.3 Constants of motion

The metric (5) does not depend on $t$, $\theta$ and $\phi$, thus the metric is simply statically spherically symmetric as it depends only on $r$ coordinate.

In terms of Lagrangian mechanics ($\mathcal{L} = \frac{1}{2} g_{\mu\nu} \dot{x}^\mu \dot{x}^\nu$), the $t$ and $\phi$ coordinates are cyclic which leads to two conserved quantities namely energy $\varepsilon$ and angular momentum $\ell_z$. For the above defined Lagrangian the corresponding conserved quantities can be calculated using the Lagrangian equation.

We can calculate the conserved quantities corresponding to cyclic coordinates $t$ and $\phi$ as $\frac{d}{d\tau}\frac{\partial \mathcal{L}}{\partial \dot{t}} = 0$, and $\frac{d}{d\tau}\frac{\partial \mathcal{L}}{\partial \dot{\phi}} = 0$, yielding $\frac{\partial \mathcal{L}}{\partial \dot{t}} = \mathcal{E} \equiv -p_\mu \xi^\mu_{(t)}/m$, and $\frac{\partial \mathcal{L}}{\partial \dot{\phi}} = L_z \equiv p_\mu \xi^\mu_{(\phi)}/m$. Constants of motion for a neutral particle in the absence of magnetic field are;

$$\dot{t} = \frac{\varepsilon}{f(r)}, \quad \dot{\phi} = -\frac{\ell_z}{r^2 \sin^2\theta}. \qquad (9)$$

From the normalization condition, $g_{\mu\nu} u^\mu u^\mu = -1$, we have the equation for $\dot{r}$,

$$u^\mu = (\frac{dt}{d\tau}, \frac{dr}{d\tau}, \frac{d\theta}{d\tau}, \frac{d\phi}{d\tau}), \tag{10}$$

so we get,

$$\varepsilon^2 - \dot{r}^2 = \left(1 + \frac{\ell^2}{r^2}\right)f(r) \implies \varepsilon^2 - \dot{r}^2 = V_{eff}(r, \varepsilon, \ell_z). \tag{11}$$

Here we are doing the calculation for equatorial plane ($\theta = \frac{\pi}{2}$) $\implies \dot{\theta} = 0$ and $V_{eff}$ (the effective potential) for the considered geometry of BH,

$$V_{\text{eff}}(r, \varepsilon, \ell_z) = \left(1 + \frac{\ell^2}{r^2}\right)f(r). \tag{12}$$

Equation (12) is the effective potential of the BH which is asymptotically flat ($V_{\text{eff}} \to 1$ as $r \to \infty$).

The Lagrangian of the particle of mass $m$ and charge $q$ moving in an external magnetic field of a curved space-time is given (see [23] and references given therein)

$$\mathcal{L} = \frac{1}{2}g_{\mu\nu}\dot{x}^\mu\dot{x}^\nu + \frac{qA_\mu}{m}\dot{x}^\mu, \tag{13}$$

and generalized 4-momentum of the particle is, $P_\mu = mu_\mu + qA_\mu$.

Now the conserved quantities (constant of motion) for charged particle will remain conserved with little modification due to magnetic field,

$$\varepsilon_B = \varepsilon = f(r)\dot{t}$$
$$\ell_B = = -r^2 \sin^2\theta \ \dot{\phi} + b, \quad b = \frac{q\mathcal{B}}{2m}. \tag{14}$$

The constants of motion regarding the charged particle in the presence of magnetic field are: particle's energy ($\varepsilon_B$) and angular momentum ($\ell_B$) given in (14).

Using the normalization condition $u^\mu u_\nu = -1$ along with Eq. (14), we can calculate the equation for $\dot{r}$ and effective potential $V_{eff}$. So the effective potential and the radial equation after taking account of the effects of magnetic field become,

$$\varepsilon^2 - \dot{r}^2 = \left(1 + r^2\left(b - \frac{\ell_z}{r^2}\right)^2\right)f(r) \implies \varepsilon^2 - \dot{r}^2 = V_{\text{eff}}(r, \varepsilon, \ell_z). \tag{15}$$

Here we are doing the calculation for equatorial plane ($\theta = \frac{\pi}{2}$) $\implies \dot{\theta} = 0$ and effective potential $V_{\text{eff}}$ in the presence of magnetic field is,

$$V_{\text{eff}}(r, \varepsilon, \ell_z) = \left(1 + r^2\left(b - \frac{\ell_z}{r^2}\right)^2\right)f(r). \tag{16}$$

Equation (16) represents the effective potential of BH in the presence of magnetic field (b). It also depends on the behavior of magnetic field with respect to the distance from the BH. There are many models of active galactic nuclei (AGN) containing a massive black hole surrounded by a magnetized accretion disc [27].

The effective force on a particle moving in the surrounding of BH is defined as,

$$F = -\frac{1}{2}\nabla_r(V_{\text{eff}}(r, \varepsilon, \ell_z)), \quad \nabla_r = \frac{\partial}{\partial r}. \tag{17}$$

Similarly we can write for charged particle,

$$F = -\frac{1}{2}\nabla_r(V_{\text{eff}}(r, \varepsilon_b, \ell_b)). \tag{18}$$

$$F = \frac{3l^2(4M - r)}{r^5} + \frac{2M(1 - 2\mathcal{B}\ell)}{r^3}$$
$$+ \mathcal{B}^2 + \frac{2\mathcal{B}(l - 3\mathcal{B}r^2) - 1}{a^2} + \eta^2\left(\mathcal{B}^2 + \frac{3\ell^2}{r^4}\right) \tag{19}$$

The effective force on the particle is repulsive because of the specific metric (S-AdS) given in Eq. (5) with DE parameter $a$ not equal to zero. The parametric force Eq. (19) can be discussed term by term as it is done in [25] and [35]. The first and second terms have $1/r^5$ and $1/r^3$ dependence, thus they are important only near the horizon. Third term of Eq. (19) represent a constant force due to magnetic field as we have a static axisymmetric magnetic field. The fourth term shows the force due to dark energy parameter $a$ and this term will remain positive if it satisfies the following condition $\ell/\mathcal{B} > 1/2\mathcal{B}^2 + 3r^2$. The fifth term is always positive, although very small due to $\eta^2$ dependence. Hence the force due to DE and topological defect parameters $a$ and $\eta$ is repulsive since the forth and fifth terms are positive.

It can be seen from Fig. 3 that higher values of the parameters $a$ and $b$ make the effective force more repulsive. Hence, the presence of DE and magnetic field shifts the position of the stable orbits away from the EH.

### 3.4 Dynamics of charged particle in the vicinity of magnetized BH

Here we have investigated the trajectories of the charged and neutral particles moving in the surrounding of magnetized BH.

Our main focus is to examine the properties of marginally stable circular orbits for test particles moving on the equatorial plane of the considered BH geometry. Such trajectories are called ISCO [26] and [37]. Further we are discussing the circular orbit of specific radius $r_0$, both the initial radial velocity ($\frac{dr}{d\tau} = 0$) and the radial acceleration ($\frac{d^2r}{d\tau^2} = 0$) must vanish for circular orbits, therefore,

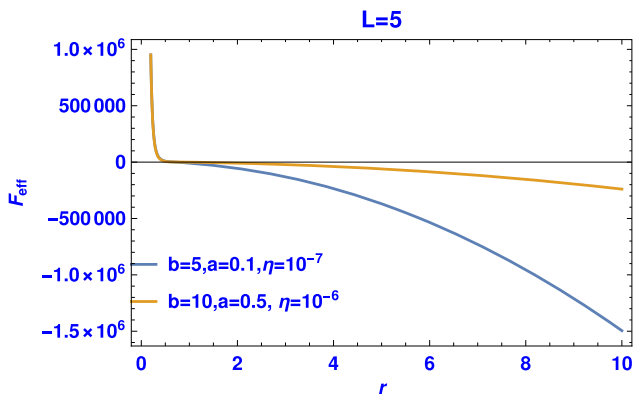

**Fig. 3** This figure shows the effective force on the particle moving in the surrounding of BH with respect to the distance ($r = r/M$) from the BH

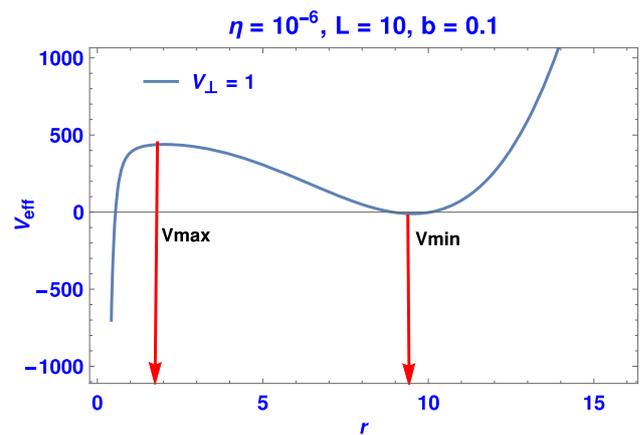

**Fig. 4** Here we plotted the effective potential against $r$. The local minimum ($V_{min}$) and maximum ($V_{max}$) are corresponds to ISCO and unstable orbits, respectively

$$V_{\text{eff}}(r, \varepsilon, \ell_z) = \varepsilon^2 \quad (20)$$

To get the stability condition for the orbits we will use the first and second derivative test as follow:

$$\left(\frac{\partial V_{\text{eff}}(r, \varepsilon, \ell_z)}{\partial r}\right)\bigg|_{r=r_0} = 0, \quad \left(\frac{\partial^2 V_{\text{eff}}(r, \varepsilon, \ell_z)}{\partial r^2}\right)\bigg|_{r=r_0} \geq 0. \quad (21)$$

One can solve Eqs. (20–21) for the radius of ISCO and for critical values of magnetic field $b$ and angular momentum $\ell_z$ at ISCO. But the equation $\frac{\partial V_{\text{eff}}}{\partial r} = 0$ is a polynomial of 7th order so, we can solve it only numerically and it has both real and imaginary roots.

Consider the particle in the ISCO which collides with another incoming particle. After collision between these particles, three cases are possible for the motion of the particles: (i) bounded motion, (ii) capture by the black hole or (iii) escape to infinity. The result depends on the collision process. For a small change in energy and momentum, the orbit of the particle will be slightly perturbed. After the collision the particle has new values of energy and momentum $\varepsilon, \ell_z$, respectively. Moreover, the total angular momentum is defined as:

$$\ell^2 = r^4(\dot\theta^2 + \sin^2\theta\,\dot\phi^2) = r^2 v_\perp^2 + r^4 \sin^2\theta \dot\phi^2, \quad v_\perp = r\dot\theta. \quad (22)$$

Now we consider the change in energy and angular momentum of the particle after the collision. We incorporate these changes by replacing the azimuthal angular momentum $\ell_z$ with total angular momentum $L$ and effective potential becomes $V_{eff}(r, \varepsilon, \ell_z) \to V_{eff}(r, \varepsilon, \ell)$. Physically it means that the particle gains some energy after the collision. So, the new energy or effective potential after collision is

$$V_{eff}(r, \varepsilon, \ell) = f(r)\left(\frac{(r(v_\perp - 2br) + l)^2}{r^2} - 1\right). \quad (23)$$

The new effective potential $V_{eff}(r, \varepsilon, \ell)$ is given by Eq. 23.

Hence the minimum energy required for a particle to escape is $V_{eff}(r, \varepsilon, \ell) \geq 1$. All the orbits with $V_{eff}(r, \varepsilon, \ell) \geq 1$ are unbounded in the sense that the particle escapes to infinity. Conversely for $V_{eff}(r, \varepsilon, \ell) < 1$, the particle cannot escape to infinity (the orbits are bounded).

Behavior of the effective potential against different parameters is presented in Fig. 5. One can see from panel (a) of Fig. 5, for large value of $b$ effective potential is very high, indicating that the possibility of a particle to fall into the BH is small. So the presence of magnetic field increase the escape possibilities of particles from the vicinity of BH. Panel (b) of Fig. 5 shows the behavior of effective potential against different values of $L$, local minimum value of the effective potential which corresponds to ISCO position shifts away from the EH for large $L$. Hence the orbits are more stable for small $b$ and large $L$, and less stable for large $b$ and small $L$.

New aspects regarding the dynamics of particle arise due to the parameters of topological defects and DE that make this regime interesting as it can lead to qualitatively new features for circular orbits. We plot the effective potential for different value of parameters $\eta$ and $a$ in panel (c) and (d). The scale of symmetry breaking parameter $\eta$ has a typical value $\sim 10^{-6}$. For an increase of one order of magnitude in the value of parameter $\eta \approx 10^{-5}$ effective potential shoots up to very large value which indicates an easy escape. But for smaller values of $\eta \approx 10^{-6}$ or 0 the effective potential $\to 0$; so particle can easily fall into the BH. Mass of the BH increases if number of particles falling into the BH increases with the condition on the value of $\eta \approx 10^{-6}$, mass of the BH might increase. Local minimum of the potential moves away from the EH with the consideration of DE parameter ($\lambda = 3/a^2$). DE is a repulsive gravitational force so it can shift the position of the ISCO. Thus stability of the orbits decreases in the presence of DE that leads to chaotic motion. Therefore, due to repulsive force of DE the size of the BH-

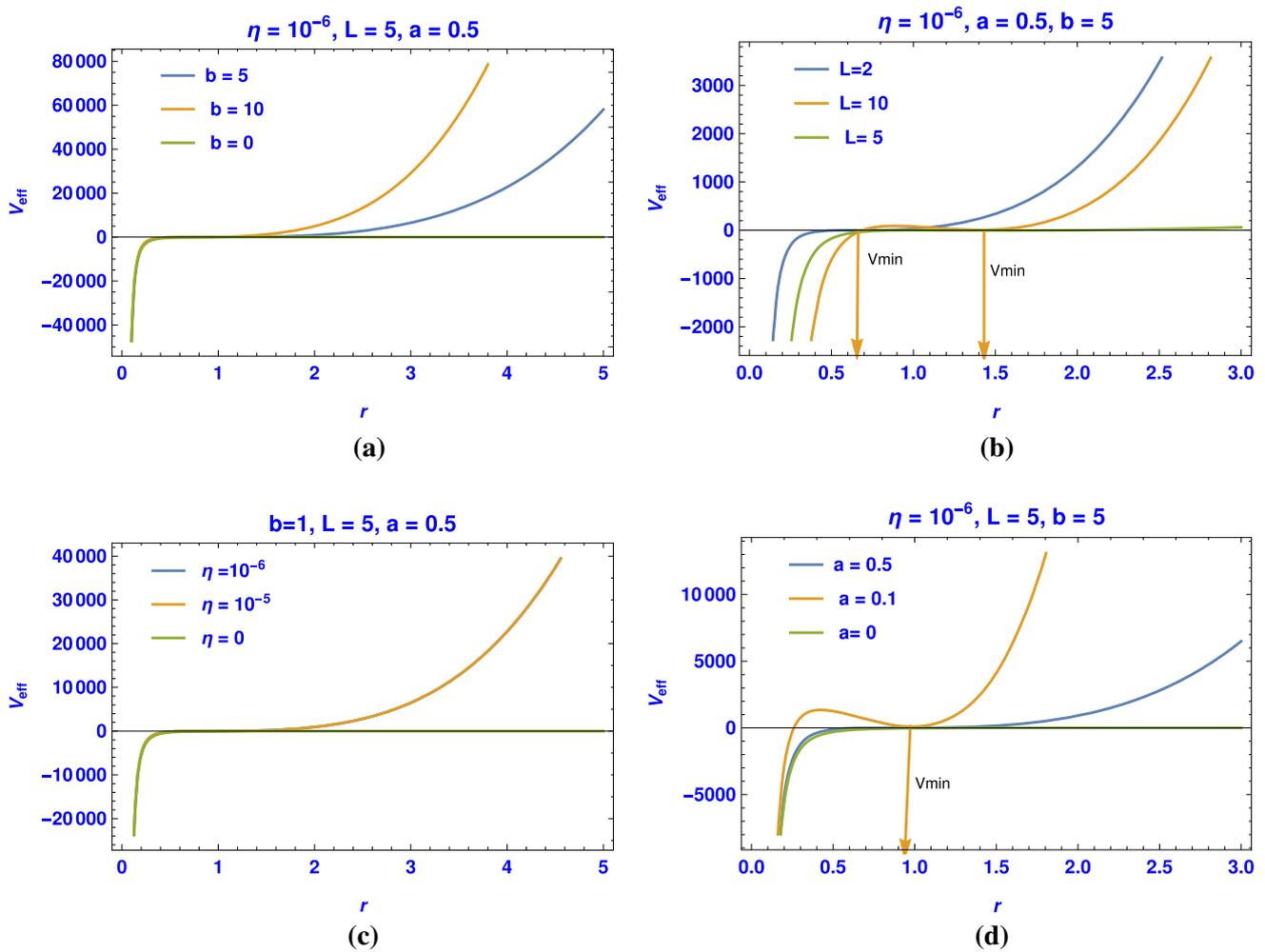

**Fig. 5** This figure represents the behavior of effective potential $V_{\text{eff}}$ against different parameters of the BH. The local minimum value $V_{\min}$ of the $V_{\text{eff}}$ corresponds to the ISCO position of the particle. **a** Shows that $V_{\text{eff}}$ is large for greater value of magnetic field; which indicates the easy escape of the particle from the BH surrounding. **b** Shows the shift in ISCO position away from the BH horizon for large value of angular momentum $L$. The behavior of $V_{\text{eff}}$ against parameter $\eta$ and $a$ are presented in **c** and **d**, respectively

EH increases. All the parameters we used in the plots are in dimensionless form ($r = r/M$, $L = \ell/M$ and b).

One can see the local minima and maxima of the effective potential in Fig. 4. This local minima corresponds to ISCO and maxima is for unstable orbits. There are two possibilities for a particle in an unstable orbit either captured by the BH or escape to infinity.

Further we are discussing the change in magnetic field with the ISCO position. The Eq. (24) shows the critical value of magnetic field as function of ISCO radius $r_o$.

In Fig. 6 we show the critical behavior of magnetic field $b$ and angular momentum $L$ as a function of ISCO radius $r_o$. So, for large $r$, $b$ becomes constant but $L$ is increasing see Fig. 6. For the weakly magnetized BH magnetic field strength decreases with the increase of distance. Panle (b) of Fig. 6 indicates that the large $L$ can shift the ISCO position $r_o$ away from the EH.

$$\dot{\phi} = \frac{\ell - b}{r^2} \implies T = \frac{2\pi r^2}{\ell - b} \qquad (25)$$

$$b_o = \sqrt{\frac{0.65 r_o^8 + 0.018 r_o^5 - 0.0046 r_o^2 - 0.5\sqrt{r_o^4 \left(r_o^3 + 0.1\right)^3 \left(r_o^6 + 0.68 r_o^3 - 0.031\right)}}{r_o^4 \left(r_o^9 - 1.05 r_o^6 + 0.3 r_o^3 - 0.0125\right)}} \qquad (24)$$

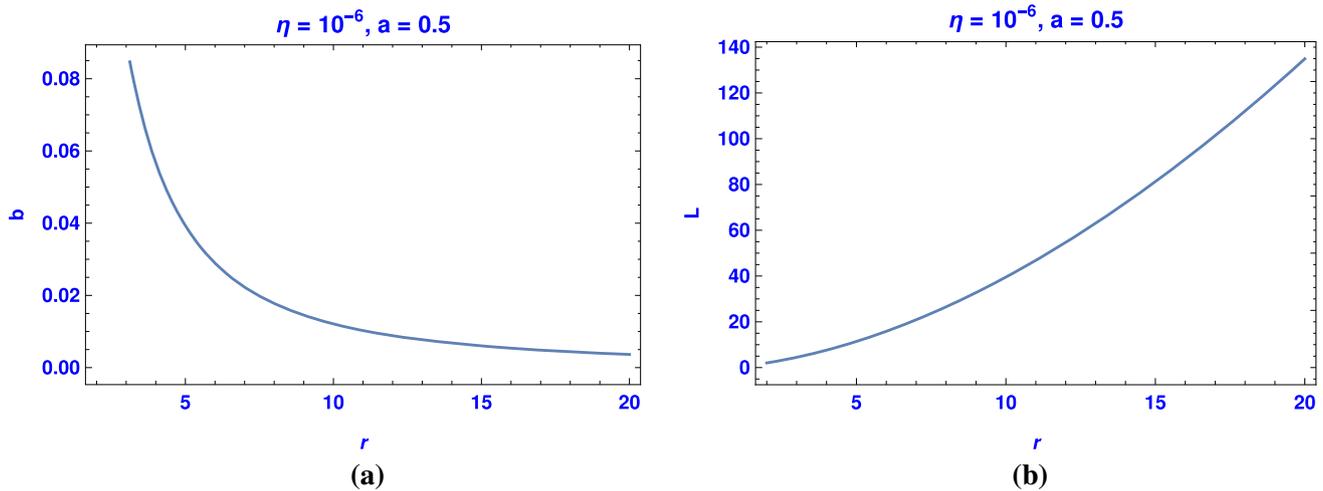

**Fig. 6** Here we have plotted the critical values of angular momentum and magnetic field at ISCO positions

The Eq. (25) shows the time of a particle moving in an orbit around a BH. In general, time to stay in an orbit around the BH is finite. The time will become infinite if angular momentum is equal to magnetic field, $\ell = b$. In the absence of magnetic field $T$ becomes infinite for $\ell = 0$. If the angular momentum is zero then orbits becomes static. It is very less likely that $\ell$ becomes equal to $b$ even if the radius of the orbit is fixed.

### 3.5 Escape velocity

According to the Blandford and Znajek [27,28] mechanism charged particles can be accelerated up to a large distance from an active galactic nuclei (AGN). After collision the particle may acquire an escape velocity ($v_\perp$) in an orthogonal direction of the equatorial plane [36]. We calculate the critical escape velocity required for an orbiting charged particle to escape to infinity.

geneous magnetic field. This type of motion is studied in detail [22]. We would like to know whether the particle will escape to infinity or not. To simplify the problem, we consider a particle initially moving in an ISCO. During the motion, the particle stays in the vicinity of the equatorial plane, crosses the plane many times and forms a compact cloud in the corresponding phase space [23]. However, such a particle eventually falls into the black hole or escapes to infinity. We study the escape velocity of the charged particle with respect to different parameters for the strength of magnetic field and angular momentum. It can be seen from Fig. 7 that for large angular momentum and magnetic field it is easy for a particle to escape and vice versa. The effect of DE parameter on the escape velocity is significant only near the BH as shown in panel (b) of Fig. 7 and it does not effect the trajectories of the particles at large distance. Motion of the particle near the horizon is chaotic because the particle is under the influence of gravitomagnetic forces and DE also play its part. For dif-

$$V_{escape} \geq \frac{a^2\left(2M + \left(\eta^2 - 2\right)r\right) - r^3}{\sqrt{\left(a^2\left(-2M + \eta^2(-r) + r\right) + r^3\right)\left(r^3 - a^2\left(2M + \left(\eta^2 - 2\right)r\right)\right)}} + 2br - \frac{l}{r} \quad (26)$$

The Eq. (26) is obtained using the Eq. (23). The equality in the above Eq. (26) corresponds to $V_{eff}(r, \varepsilon, \ell) = 1$. Therefore, the escape condition for the particle from the BH vicinity is given by the Eq. (26).

For small values of the transferred energy and momentum the orbit will be only slightly perturbed. However, for larger values of $\epsilon - \epsilon_0$ the particle can go away from the initial plane, and finally be captured by the BH or escape to infinity.

Suppose a kicked particle escapes to infinity where the gravitational field of the BH vanishes. Hence, such a charged particle is moving in a practically flat spacetime with a homo-ferent value of parameter $a$ the escape velocity behavior is same but near the horizon it is chaotic.

In Fig. 8 we plotted the escape velocity of particles moving in the co-rotating and counter-rotating orbits around the BH. Difference in their escape velocity is prominent only near the BH and it looks similar for large distance from the EH. The shaded region in Fig. 8 corresponds to the escape velocity of the particle and the solid curve represents the minimum velocity required to escape from the surrounding of the EH to infinity. The unshaded region represents the bound motion around the BH. It can be seen from Fig. 8 that the escape

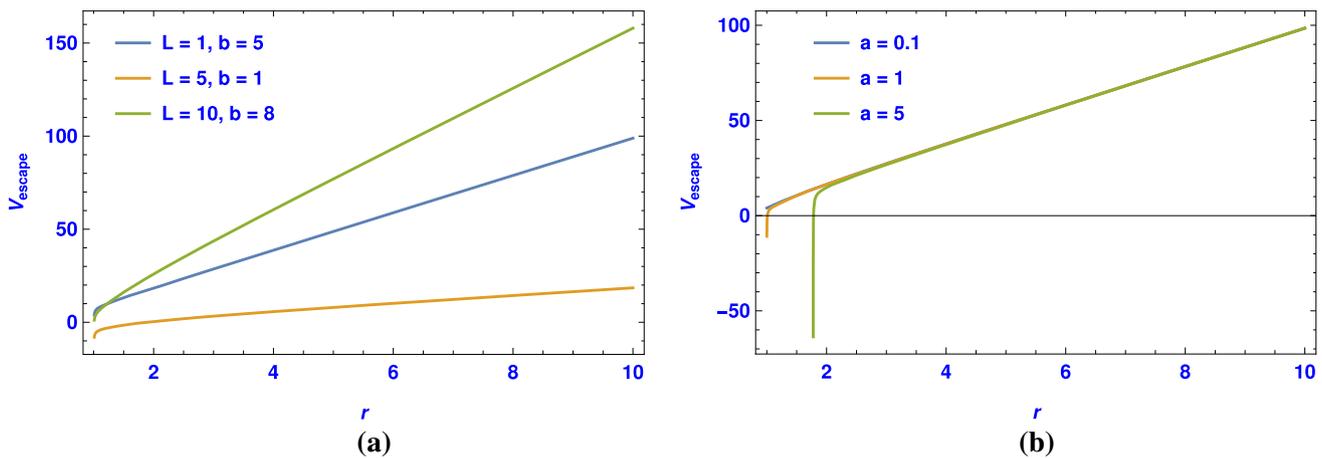

**Fig. 7** In **a** we have plotted the escape velocity of the orbiting particles for different values of b and L which represent that the higher the value of L and b, it is easy for a particle to escape. **b** Shows the behavior of escape velocity against $r$ for different value of DE parameter $a$. So, the DE effects the trajectories of the particles only near the horizon shown in (**b**)

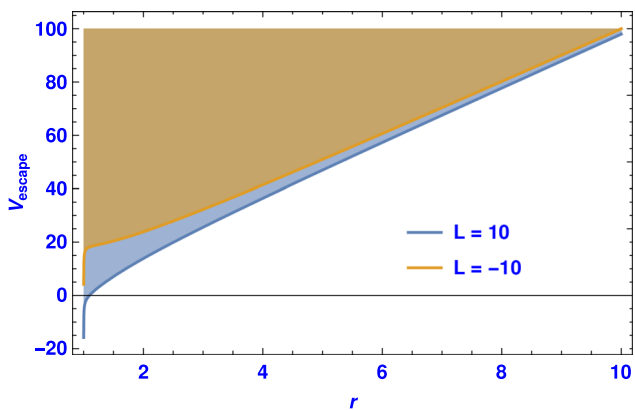

**Fig. 8** In this figure we have plotted the escape velocity for co-rotating ($L = 10$) and counter-rotating ($L = -10$) particles around the BH with parameters $a = 1$ and $b = 5$. Shaded and unshaded regions correspond to escape and bound motion around the BH, respectively

possibilities for a counter-rotating particle are more in comparison to the co-rotating particles.

## 4 Results and discussion

Here we have discussed the ability of a magnetized BH to work as a particle accelerator. We explored that how a weak magnetic field can make the orbits more stable. We determined the motion of the charged particle using an effective potential approach. Collision of two particles is demonstrated, one of these is charged and revolving at ISCO and the other is neutral and falling from infinity. There exist two different types of behavior for the particle moving in ISCO: captured by BH, and escape to spatial infinity, $z = \pm\infty$. When the charged particle is moving for a long time in the vicinity of magnetized BH close to its equatorial plane, crossing it again and again, it may gain enough energy during each crossing needed to escape. This kind of motion has features similar to a diffusion process in different turbulent environments. The main effect on the motion of the particle near the BH is due to the gravitational force of the BH. But, far from the EH, electromagnetic force is the main agent to determine the motion of the particle.

Effect of different parameters on the effective potential of BH with respect to the dynamics of the particle is shown in Fig. 5. In the absence of magnetic field particles coming from infinity have to face small potential barrier to fall into BH than its presence. Hence, in the presence of magnetic field the stable circular orbits are more likely to exist in comparison to its absence. So, for any $V_{\text{eff}} > 0$ it is possible for a magnetic field to hold plasma in equilibrium outside the EH relative to a BH and this equilibrium need not be subject to the kind of relativistic instability that destroys circular orbits. Therefore, we have found that the possibility for a particle to fall into BH decreases in the presence of magnetic field and it can provide enough energy to the diffused particle near the BH needed to escape.

Angular momentum of the particle also plays an important role regarding the motion of the particle. Local minimum of the effective potential which corresponds to ISCO position shifts away from the EH for large value of angular momentum. The closeness of ISCO to the horizon is controlled by the value of the magnetic field and angular momentum. We have found a significant shift in the ISCO position for large value of angular momentum and magnetic field. Particles with small angular momentum and magnetic field can be captured easily by the BH gravity.

We have computed the escape velocity for the charged particle moving in the ISCO and discussed it for co-rotating and counter-rotating orbits. It is found that the counter-rotating particles can escape easily from the BH vicinity than co-rotating. It is also found that if the angular momentum of the particle is large and the strength of magnetic field is high then it can escape easily. The effect of DE on the escape velocity is dominant only near the EH.

We have also discussed the positioning of ISCO and size of EH due to the DE $a$ and symmetry breaking $\eta$ parameters. The possibilities for a particle to fall into the BH are increased due to topological defect. We also explored the increase of EH of BH with the increase of DE, $\lambda = 3/a$ which indicated the increase of mass of BH. We calculate the effective force on the charged particle while moving in the vicinity of BH and also inferred the conditions when the force on the particle due to dark energy is attractive and when it is repulsive. DE is basically a repulsive force, it can shift the position of ISCO away from the EH. We have found that the AH of the BH increases due to DE which is also an indication of increase in the mass of BH. Regarding the mass of the BH there are two different conclusions that the BH mass decreases due to the accretion of phantom energy [18] and it increases due to the same reason [19]. Our results match with the later [19].

If the BH mass increases, and the future universe is dominated by phantom DE, then the BH-AH and the cosmic-AH will eventually coincide, after which both horizons disappear and the singularity becomes naked. It happens in finite co-moving time before the Big Rip occurs, violating the Cosmic Censorship Conjecture. Further more, DE is not constant rather growing density [3] and hence stronger over the cosmic time. If the predicted phantom DE model is correct then the push from phantom energy would grow without bounds, finally overcoming gravity, thus Big Rip is inevitable.

Mass of the BH increases if more matter falls into the BH. This phenomenon is a prototype of accretion of matter onto the BH. Here we considered the effect of gravity and electromagnetic force and DE on the charged particles moving around the BH. There are many other phenomena occurring near the BH (e.g., AGN feedback, star formation and etc) which contribute significantly in the accretion process.

Moreover, we have found that the system is chaotic regarding the stability of orbits. To study the instabilities of the orbits with respect to different parameters using Lyapunov coefficient is our potential future work.

**Data Availability Statement** This manuscript has no associated data or the data will not be deposited. [Authors' comment: Because this study is not directly related to any experiment or simulation data. Moreover, this is a parametric study about the motion of particle around the black hole. All the equations with parameters are already available in the article if one wants to reproduce the results.]

Funded by SCOAP³.